\documentclass[iop]{emulateapj}
\usepackage{lscape}

\shorttitle{Mid-Infrared Search for Substellar Companions of Nearby Planet-Host Stars}
\shortauthors{Hulsebus, Marengo, Carson, Stapelfeldt}

\begin{document}

\title{A Mid-Infrared Search for Substellar Companions of Nearby Planet-Host Stars}

\author{A. Hulsebus, M. Marengo}
\affil{Dept of Physics and Astronomy, 12 Physics Hall, Iowa State University, Ames, IA 50010}

\author{J. Carson}
\affil{Department of Physics and Astronomy, College of Charleston, 58 Coming St., Charleston, SC 29424}

\author{K. Stapelfeldt}
\affil{Exoplanets and Stellar Astrophysics Laboratory,
 Code 667, NASA Goddard Space Flight Center, Greenbelt MD 20771}

\begin{abstract}

Determining the presence of widely separated substellar-mass companion is crucial to understand the dynamics of inner planets in extrasolar planetary systems (e.g. to explain their high mean eccentricity as inner planets are perturbed by the Kozai mechanism). We report the results of our \textit{Spitzer}/Infrared Array Camera (IRAC) imaging search for widely separated (10 -- 25$^{\prime\prime}$) substellar-mass companions for 14 planet-host stars within 15~pc of the Sun. Using deep 3.6 and 4.5~$\mu$m observations in subarray mode, we found one object in the field of 47~UMa with [3.6]$-$[4.5] color similar to a T5 dwarf, which is, however, unlikely to share common proper motion with 47~UMa. We also found three objects with brown-dwarf-like [3.6]$-$[4.5] color limits in the fields of GJ~86, HD~160691, and GJ~581, as well as another in the field of HD~69830 for which we have excluded common proper motion. We provide model-based upper mass limits for unseen objects around all stars in our sample, with typical sensitivity to 10~M$_{J}$ objects from a projected separation of 50 to 300 au from the parent star. We also discuss our data analysis methods for point-spread-function subtraction, image co-alignment, and artifact subtraction of IRAC subarray images.

\end{abstract}


\section{Introduction}

Large bodies in the periphery of planetary systems have the potential to wreak havoc on the orbits of inner planets through secular interactions. \citet{naoz11} posit this effect to be responsible for the migration of hot Jupiters; \citet{marzari02} describe how planet-planet scattering can result in moderately eccentric planetary orbits; and \citet{takeda05} demonstrate how Kozai-type interactions \citep{kozai62} can result in highly eccentric planet orbits, inward of a large planet or brown dwarf.

Given that conventional planet formation models, including both core accretion \citep{pollack96} and gravitational instability \citep{boss97}, describe planets forming in mostly circular, well-aligned circumstellar disks, it is important to explain why the mean eccentricity of discovered exoplanets is currently 0.19\footnote{http://exoplanet.eu, as of Oct 7, 2013}, with many of the highest eccentricities found in single-planet systems (78\% with $e>0.5$ are single planets). If secular effects involving a massive companion are responsible, these companions may be observable, if they have not yet been ejected from the system. \citet{ford08} conclude that most planet scattering instabilities should occur on timescales comparable to the planet formation timespan, but do acknowledge that triple planet systems in certain configurations could potentially be quasi-stable on timespans from $10^{6}$ to $10^{10}$ years, so it is not impossible to have planet ejection occur in an old system.

\citet{gizis01} found brown dwarf companions $>1000$~au from the primary, and \citet{luhman07} found HD~3651B, a brown dwarf orbiting 476~au from a star with an eccentric planet. However, the same formation models mentioned above do not explain how such a massive companion could form so far from the parent star: that far-separated substellar companions and highly eccentric planets both exist deserves explanation. One such explanation is that all objects form close to their parent stars and end up in their current configurations through scattering events (\citealt{boss06}, \citealt{nagasawa11}).

Whether causing havoc or not, widely separated companion brown dwarfs are interesting in their own right, as relatively few are known: of the 992\footnotemark[1] substellar-mass companion objects discovered, only 36 (all with mass $>1~M_{J}$) are farther than 10~au from their parent, and 21 of those lie between 10-300~au. In part, this is a selection effect: the timescales necessary to find these objects with either radial velocity (RV) or transit methods are prohibitive, and all of the aforementioned brown dwarfs have been detected by direct imaging. Ground-based direct imaging searches have been successful in finding gas giant companions (e.g. \citealt{carson13}, \citealt{marois10}, \citealt{chauvin05} and \citealt{neuhauser05}), but have lower sensitivity to the coolest T- and Y-dwarfs (which have distinctive [3.6]$-$[4.5] colors) because of telluric absorption in the thermal infrared. Space telescopes get around this problem, and WISE has been successful in finding many field T- and Y-dwarfs \citep{kirkpatrick11}, but is less sensitive to brown dwarfs around stars because of the high contrast ratios involved. Studies like \citealt{bergfors13}, \citealt{ginski12}, and \citealt{roell12} have searched for stellar-mass companions to planet-host stars, but are generally not sensitive to brown-dwarf-mass objects.

The inner working angle of any direct-imaging companion search is limited by the brightness of the parent star compared to its companion. At optical wavelengths, this is determined by the contrast ratio between the parent star's emission and the companion's reflected light. This can be problematic at close angular distances or if the starlight saturates the detector. However, the contrast ratio in the infrared (e.g. at 3.6~$\mu$m), where thermal emission dominates over reflected light, is much more manageable. This lower ratio opens up the possibility of searching for planetary mass companions with advanced Point Spread Function (PSF) subtraction techniques, provided that the PSF is both stable and well spatially sampled. For these reasons, we have used the InfraRed Array Camera (IRAC, \citealt{fazio04}) onboard the Spitzer Space Telescope \citep{werner04} to search for companions around 14 nearby stars. Using IRAC gives the added advantage of being able to characterize T-dwarfs, in particular, using their characteristically red [3.6]-[4.5] color due to methane absorption bands \citep{burrows03}.

This work builds on our previous IRAC low-mass companion searches (e.g. \citealt{marengo06}, \citeyear{marengo09}, \citealt{carson11}), which have resulted in the discovery of two brown dwarf companions \citep{luhman07}, including a T7.5 companion of the exoplanet host star HD~3651. In \citet{marengo09}, in particular, we have developed the same technique used in the work presented here, adopting IRAC shorter frame-time subarray observing mode. This allows us to reduce the primary star saturation and the area with high PSF-subtraction residuals, narrowing our inner working angle from 20$^{\prime\prime}$ to 5$^{\prime\prime}$. Building on this previous work, we used IRAC's subarray mode to probe for $>5~M_{J}$ companions at separations of 25 to 350~au (for a star 15~pc distant) around a sample of 14 nearby planet host stars.

Herein, we present our target list in Section~\ref{targetselection}, our analysis methods (Section~\ref{dataanalysis}), sensitivity limits (Section~\ref{sensitivitysection}), and companion candidates selected on the basis of their IRAC colors (Section~\ref{detectedsources}). In Section~\ref{discussion} we discuss our results.

\section{Target selection and Observations}\label{targetselection}

Because so few brown dwarf companions to exoplanetary systems are known, it would be equally interesting to find a perturbing brown dwarf companion as to find one around an exoplanet system that does not exhibit abnormal eccentricity, or likewise one that is bound to a system known to already have another brown dwarf companion. Therefore, we selected 14 target stars to observe (Table~\ref{targetstardata}) with only the criteria that they be located within $\sim$15~pc from the sun and have exoplanets detected by radial velocity. This 15~pc distance restriction was chosen to provide the best inner working angle, equivalent to a few tens of au from the target star, and an outer working angle of a few hundreds of au (due to the subarray field of view): a range in which perturbing companions acting through the Kozai mechanism are expected. Target stars range in age from 0.1 to $>10$~Gyr, estimated from various factors including rotational periods, magnetic activity, galactic velocity, and emission lines (for more information, see individual notes at the bottom of Table~\ref{targetstardata}).

\begin{deluxetable*}{lcccl}[!htbp]
\tabletypesize{\scriptsize}
\tablecaption{Target star data\label{targetstardata}}
\tablehead{
\colhead{Target} & \colhead{Dist(pc)} & \colhead{Age(Gyr)} & \colhead{Sp. Type} & \colhead{Planet eccentricities (closest to farthest) and notes}
}
\startdata
    SCR~J1845-6357	& $3.85\pm0.02$\tablenotemark{ab}  & $2.45\pm0.65$\tablenotemark{c}	& M8.5V\tablenotemark{ab}  & $45\pm20$~M$_{J}$ T6 companion closer than 4.5~au\tablenotemark{aj}\\
    GJ~876 			& $4.69\pm0.05$\tablenotemark{z}  & $2.5\pm2.4$\tablenotemark{d}  & M5.0V\tablenotemark{ad}	 & 0.207$\pm$0.055, 0.25591$\pm$0.00003, 0.0324$\pm$0.0013, 0.055$\pm$0.012\tablenotemark{e}\\
    GJ~581 			& $\sim6.3$\tablenotemark{ae}  & $8.5\pm1.5$\tablenotemark{f}	& M5V\tablenotemark{ad} 	 & 0.031$\pm$0.014, 0.07$\pm$0.06, 0.205$\pm$0.08, 0.32$\pm$0.09\tablenotemark{g}\\
    GJ~849 			& $9.09\pm0.1$\tablenotemark{z}  & $>3$\tablenotemark{h}	& M3.5V\tablenotemark{ad} 	& 0.04$\pm$0.02\tablenotemark{i}\\
    GJ~436 			& $\sim10.2$\tablenotemark{ae} & $6\pm5$\tablenotemark{j}	& M3.5V\tablenotemark{ad} 	& 0.15$\pm$0.012\tablenotemark{k}\\
    GJ~86 			& $10.78\pm0.04$\tablenotemark{z} & $2.94^{+9.56}_{-1.93}$\tablenotemark{l}	& G9V\tablenotemark{af}  & 0.046$\pm$0.004, 1.7$^{\prime\prime}$ WD binary\tablenotemark{m}\\
    HD~3651 		& $11.06\pm0.04$\tablenotemark{z} & $4.41\pm3.29$\tablenotemark{l}	& K0V\tablenotemark{ag} 	& 0.63$\pm$0.04, T-dwarf companion\tablenotemark{n}\tablenotemark{al}\\
    55~Cnc 			& $12.34\pm0.11$\tablenotemark{z} & $10.2\pm2.5$\tablenotemark{q} & G8V\tablenotemark{aa} 	& $>0.06$, 0.0159$\pm$0.008, 0.053, 0.0002, 0.025$\pm$0.03\tablenotemark{r}\tablenotemark{s}\\
    HD~69830 		& $12.49\pm0.05$\tablenotemark{z} & $7\pm3$\tablenotemark{o} 	  & G8+V\tablenotemark{ah}	& 0.1$\pm$0.04, 0.13$\pm$0.06, 0.07$\pm$0.07, debris disk\tablenotemark{o}, 10$^{\prime\prime}$ candidate M companion\tablenotemark{am}\\
    HD~147513 		& $12.78\pm0.06$\tablenotemark{z} & $\sim$0.4\tablenotemark{ak}	& G1V\tablenotemark{ak}  & 0.26$\pm$0.05\tablenotemark{p}, possible 5360~au WD binary\tablenotemark{p}  \\
    ups And 		& $13.49\pm0.03$\tablenotemark{z} & $3.8\pm1.0$\tablenotemark{t}	& F9V\tablenotemark{ai}  & 0.013$\pm$0.016, 0.24, 0.274, 0.00536$\pm$0.00044\tablenotemark{u}\tablenotemark{v}\\
    47~UMa 			& $14.06\pm0.05$\tablenotemark{z} & $7.4\pm1.9$\tablenotemark{a}  & G1V\tablenotemark{aa}  & 0.032$\pm$0.014,  $0.098^{+.047}_{-.096}$,  $0.16^{+.09}_{-.16}$\tablenotemark{b}  \\
    HD~160691 		& $15.51\pm0.08$\tablenotemark{z} & $5.17\pm4.45$\tablenotemark{l} & G3IV-V\tablenotemark{ah}	& 0.172$\pm$0.04, 0.0666$\pm$0.0122, 0.128$\pm$0.017, 0.0985$\pm$0.0627\tablenotemark{x}\\
    51 Peg 			& $15.61\pm0.09$\tablenotemark{z} & $4.0\pm2.5$\tablenotemark{a}	& G2.5IVa\tablenotemark{aa} 	& hot Jupiter with $e<0.01$\tablenotemark{w}\\
\enddata
\tablecomments{
(a) \citealt{fuhrmann97} (b) \citealt{gregory10} (c) \citealt{kasper07} (d) \citealt{correia10} (e) \citealt{rivera10} (f) \citealt{selsis07} (g) \citealt{forveille11} (h) \citealt{butler06} (i) \citealt{bonfils12} (j) \citealt{torres07} (k) \citealt{deming07} (l) \citealt{saffe05} (m) \citealt{queloz00} (n) \citealt{fischer03} (o) \citealt{lovis06} (p) \citealt{mayor04} (q) \citealt{vonbraun11} (r) \citealt{demory12} (s) \citealt{fischer08} (t) \citealt{fuhrmann98} (u) \citealt{curiel10} (v) \citealt{barnes11} (w) \citealt{marcy97} (x) \citealt{pepe07} (y) \citealt{2mass} (z) \citealt{vanleeuwen07} (aa) \citealt{montes01}
(ab) \citealt{deacon05} (ac) \citealt{faherty09} (ad) \citealt{jenkins09} (ae) \citealt{ESA97} (af) \citealt{torres06} (ag) \citealt{vanbelle09} (ah) \citealt{gray06} (ai) \citealt{abt09} (aj) \citealt{biller06} (ak) \citealt{mamajek08} (al) \citealt{luhman07} (am) \citealt{tanner10}
}
\end{deluxetable*}

Among the selected targets, there are 5~M, 1~K, 7~G, and 1~F stars. All are hosts to planets detected by radial velocity (eccentricities listed in Table~\ref{targetstardata}), with the exception of GJ~436, whose planet was discovered by transit. HD~69830 also has a debris disk \citep{beichman05} and candidate M-dwarf companion \citep{tanner10}. GJ~86 is binary with a white dwarf ($\sim$2$^{\prime\prime}$; \citealt{mugrauer05}), and SCR~J1845-6357, itself a brown dwarf, has a T-dwarf companion detected by direct imaging at a separation of $\sim$1.2$^{\prime\prime}$ \citep{biller06}. Both companions are below the resolving power of IRAC. HD~147513 is also binary with a white dwarf \citep{mayor04}, which, at a projected separation of 5360~au, is outside our field of view. Nine of the targets have at least one planet with $e>0.1$, which may be an indication of the presence of a perturbing companion. GJ~876 has two high-eccentricity planets inward of a 2.3~M$_{J}$ planet, which may be an example of the type of secular interactions we are exploring \citep{lee02}.

Included in the sample is HD~3651, whose planet has the highest eccentricity of our sample (e=0.63, \citealt{fischer03}), and around which we previously discovered a T-dwarf companion at a projected separation of 480~au (43$^{\prime\prime}$, see \citealt{luhman07}). This system may be an example of the type of secular interactions we seek, although \citet{anglada10} refit the RV data of HD~3651 and claim that instead of one high-eccentricity planet, there are likely two lower-eccentricity (e=0.06$\pm$0.20 and e=0.04$\pm$0.20) planets in a 2:1 period commensurability.

Similarly, $\upsilon$~And~c and d have high eccentricities (e=0.24, e=0.274, \citealt{barnes11}), which \citet{curiel10} explains as due to a 3:1 resonance between $\upsilon$~And~d and e. $\upsilon$~And~e has a mostly circular orbit, so an external perturber may not be necessary to explain the dynamics of this system.

Finally, GJ~436 has one planet with moderate eccentricity (e=0.15, \citealt{deming07}). This eccentricity was revalidated by \citet{wang11} and attributed by \citet{maness07} to an unseen 22.5M$_{E}$ object 0.0285~au from the star. If that object exists, we would not be able to resolve it.

Ten of our targets were previously observed with IRAC \citep{patten05} to search for brown dwarf companions at large separation (projected separation of $\sim$100 to $\sim$1500 au). As these observations had frame times of 30~sec, they heavily saturated the primary star, limiting our inner working angle to 20$^{\prime\prime}$ from the target (see \citealt{marengo09}). For this reason we chose to observe our target stars in subarray mode, allowing for dramatically shorter exposure times and restricting saturation to the innermost one or two pixels of the brightest sources.

Our observations were executed between Sept 2007 and Aug 2008 (PID~40976), and were modeled on our previous observations of $\epsilon$~Eridani and Fomalhaut (PID 30754, \citealt{marengo09}), adapted to the lower brightness of the selected targets. We selected frame times for K, G, and F stars were 0.1~seconds (0.08~sec integration time), and 0.4~sec (0.32~sec integration time) for M stars. Each subarray image is 32$\times$32 pixels, 1.22$^{\prime\prime}$/pix, and is composed of 64 exposures repeated for each dither position. We used the standard IRAC nine-point Reuleaux dither pattern to obtain non-redundant spatial sampling to build up a total integration time of 921.6~sec per target, with 5 or 20 images (320 or 1280 exposures) per pointing for the 0.4 and 0.1~sec exposures, respectively. The large number of exposures was required to increase the signal-to-noise ratio of potential companions. The total overlap area of the exposures is $44^{\prime\prime}\times44^{\prime\prime}$.

Only 3.6~$\mu$m and 4.5~$\mu$m observations were needed because T- and Y-dwarfs have characteristically red [3.6]$-$[4.5] colors, due to the presence of methane absorption bands near 3.3~$\mu$m \citep{burrows03}. The total integration time was set to allow the detection of 5~Gyr old, 5~M$_{J}$ planetary mass bodies at 10 pc from the Sun, based on the predicted photon noise from the PSF of the primary star \citep{marengo09}. This was our goal, but we had to compromise for an integration time that resulted in lower sensitivity. Once the PSF was subtracted, however, the detection limit in the innermost 5$^{\prime\prime}$ was not photon noise, but rather the residual noise from subtracting the PSF (see Figure~1 in \citealt{marengo09}). Outside of 5$^{\prime\prime}$, these residuals diminish and electronic artifacts dominate the noise profile. This issue was already encountered in \citet[see section 2.1]{marengo09}. In this work we have improved our artifact removal technique significantly in order to reach higher sensitivity.

\section{Data Analysis}\label{dataanalysis}

As the goal of this project is the detection of faint companions around bright stars, our data reduction procedure is designed to combine the individual exposures into a single high-dynamic-range final image, while suppressing the light from the primary star with an accurate PSF subtraction. Other direct imaging searches have employed the LOCI algorithm \citep{lafreniere07} or principal component analysis (\citealt{soummer12}; \citealt{amara12}) for PSF subtraction. Both of these methods use PSF segmentation to compensate for instabilities in the PSF by using a library of PSFs produced with a large number of roll angles. We have an insufficient number (only one) of roll angles to use these methods. Furthermore, the thermal stability of IRAC's PSF makes them less necessary, and electronic artifacts are the strongest source of noise, above the PSF subtraction residual noise in most of our field of view. Instead, our data reduction technique builds upon the work of \citet{marengo09}. We started from Basic Calibrated Data (BCD) images from the Spitzer Heritage Archive, processed with the IRAC Pipeline version S18.18.0. We then wrote our own custom procedure to combine the nine individual frames from each dither position into a final image, for each star, in each IRAC band. Standard mosaicing procedures are not suitable for our targets because, due to the small field of view of the subarray fields, the lack of background stars in the individual BCDs prevents accurate World Coordinate System (WCS) alignment of individual frames by the IRAC pipeline coordinate refining routine. As a consequence, mosaics made with the standard Spitzer Science Center MOPEX software \citep{MOPEX} would be blurred.

\begin{figure}[ht]
\includegraphics[width=\columnwidth]{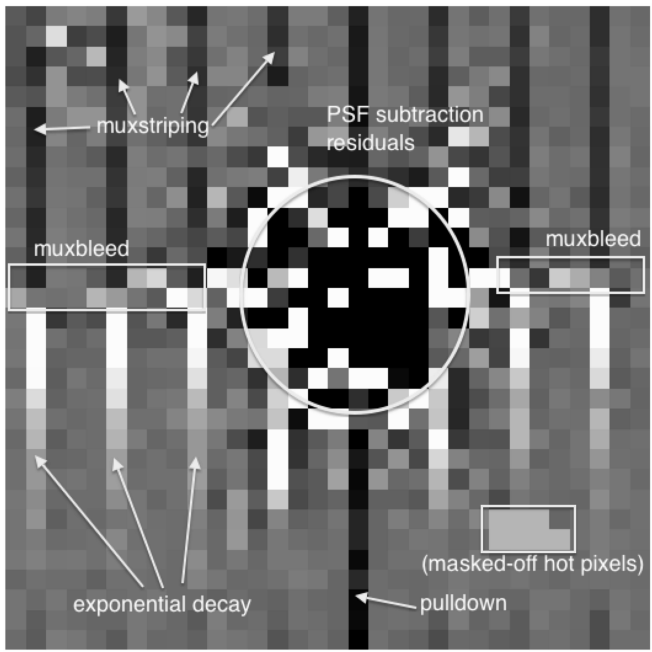}
\caption{Typical appearance of different electronic artifacts visible after PSF subtraction. PSF subtraction residuals and hot pixels are also labeled for clarity. Image shown is 47~UMa, single dither position, 3.6 $\mu$m. Field of view is $39^{\prime\prime}\times39^{\prime\prime}$.\label{electronicartifacts}}
\end{figure}

As mentioned in Section~\ref{targetselection}, for subarray, electronic artifacts dominate the noise profile of PSF-subtracted images outside of approximately $5^{\prime\prime}$ from the center of the PSF. These electronic artifacts (shown in Figure~\ref{electronicartifacts}; details provided in Section~\ref{elec}) are produced by bright stars falling on the detector array. The response of the array to bright sources differs from pixel to pixel, so these artifacts need to be characterized and removed in situ, before the nine dither positions are shifted and co-added for sub-sampling.

Two different PSF subtractions were necessary. The first was performed to remove the light from the central bright star in order to better characterize the electronic artifacts. Once characterized, the artifacts were removed from the BCDs, allowing a second, cleaner PSF subtraction to be performed to characterize the stars and search for companions. The IRAC PSF is stable, but position dependent, differing in shape across the focal plane due to light-path geometry, with the most extreme differences near the edges, where the subarray field is located. Because electronic artifact characterization requires a precise PSF subtraction at each dither position, and since there is not a reliable, high S/N ratio Spitzer IRAC PSF available for subarray mode, we created our own PSF for each star, using the other stars in our sample, on a per-dither-position basis.

\subsection{Image Alignment and First PSF Subtraction}\label{1stpsf}

\begin{figure}[ht]
\includegraphics[width=\columnwidth]{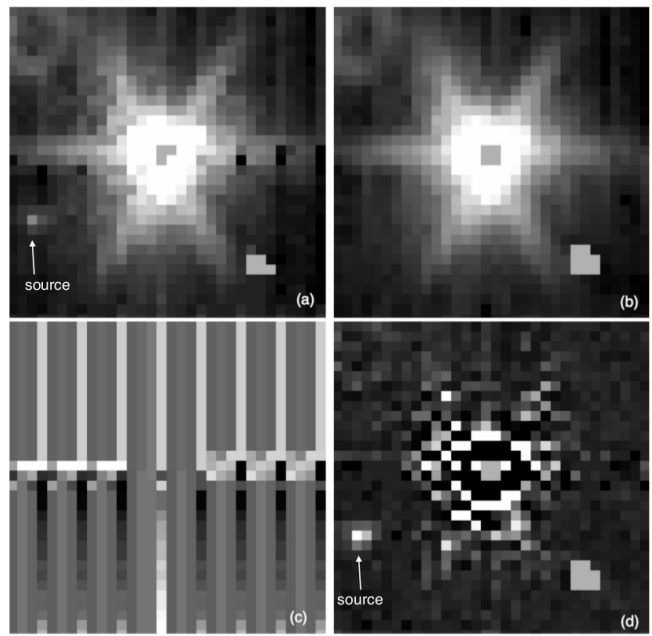}
\caption{Example of artifact removal process for GJ~581, 3.6~$\mu$m.
(a) 99\%-saturation-masked image of GJ~581, single dither position. (b) PSF created for the same dither position. (c) Combined electronic artifact correction matrix. (d) PSF- and artifact-subtracted image.
Hot pixels in the lower right of each image have been masked off. A detected source is visible to the lower left of GJ~581 in frames (a) and (d). Field of view is $39^{\prime\prime}\times39^{\prime\prime}$.
\label{process}}
\end{figure}

Figure~\ref{process} demonstrates our data reduction procedure. In order to characterize, pixel by pixel, the pattern of the electronic artifacts, we subtracted the PSF from each BCD image (Figure~\ref{process}, panel a) on the original 32$\times$32-pixel grid. In \citet{marengo09}, we used a PSF made from archival data (PID 30666), but this was not ideal due to the different dither pattern used in those observations. For this analysis, we created a PSF for each dither position from the other images in our data set. Each image was aligned and stacked at the pixel size of the original 32$\times$32 subarray grid to create a PSF (Figure~\ref{process}, panel b) for each frame.

IRAC starts to become non-linear at 10,000 and 12,000~DN in bands 1 and 2, respectively \citep{monson12}, so we masked off any pixels higher than 99\% of those threshold values. The FWHM of IRAC's 3.6~$\mu$m PSF is 1.66$^{\prime\prime}$, or 1.36 native pixels; assuming the core to be Gaussian, a 0.1-pixel alignment error affects up to 11\% of the image flux. The WCS coordinate uncertainty for our images is on the order of 1$^{\prime\prime}$, roughly 0.8 pixels, making the WCS coordinates provided with the frames too imprecise for sub-pixel alignment of the individual frames. As previously mentioned, WCS coordinates could not be refined due to a lack of field objects in the subarray images. A simple 2-D centroid of each image was also inadequate for alignment purposes: the PSF core changes shape even for small shifts of the stars on the pixel grid, due to IRAC under-sampling, pixel-phase effects, and position-dependent focus of the camera \citep{hora08}. Instead, we calculated the stationary points of the numerical gradient of the core of the PSF (its not-saturated part) along three evenly spaced (120 degrees) axes and used the intersection of those gradients as the center coordinate for each image, giving a typical alignment precision (average error) in 3.6~$\mu$m of 0.0027$^{\prime\prime}$ and 0.0028$^{\prime\prime}$ in 4.5~$\mu$m. For comparison, using a centroid for alignment of a 1.22$^{\prime\prime}$/pix image produced an average error of 0.07$^{\prime\prime}$ in 3.6~$\mu$m and 4.5~$\mu$m. $\upsilon$~And was too saturated for this technique to work and required manual adjustment.

Using these coordinates to align the images together, we created a PSF for each dither position of each star out of the other twelve star images in our data set (omitting the star itself and 55~Cnc, which has strong horizontal artifacts and was therefore not included in any PSF). We normalized each star image to the target image by finding a multiplicative scale factor and a pedestal offset between them. The scale factor was initially estimated as the ratio of the total fluxes of the two images (this would be incorrect to use as a final value because saturation and field star light differ among images), while the initial offset was set to zero. Using these initial values, we calculated the slope of the radial profile of the difference of the two images. A subtraction with perfect normalization would have zero slope, and the slope is linearly related to the scale factor. After two more scale factor ``best guesses", their respective subtraction slopes were used as the basis for a linear regression to calculate the final scale factor between the two images. Using this scale factor, we performed a similar calculation to obtain the linear offset between each image and the target. The resulting aligned, normalized frames were stacked together. Any pixels with values greater than $2\sigma$ above the median were discarded, and the mean of the remaining pixels became the PSF for that dither position of that target star (Figure~\ref{process}, panel b). By scaling our model PSF to have the same IRAC flux as Vega, we were able to use this same technique to calculate the photometry of each of our target stars (results given in Table~\ref{targetphot}).

\begin{deluxetable*}{lccccccccccc}[t]
\tabletypesize{\scriptsize}
\tablecaption{Photometry of target stars\label{targetphot}}
\tablehead{
\colhead{Star} & \colhead{[J]} & \colhead{[H]} & \colhead{[K]} & \colhead{Date Observed} & \colhead{[3.6]} & \colhead{[4.5]}
}
\startdata
SCR~J1845-6357\tablenotemark{*} & 9.54 & 8.97 & 8.51 & 2007-10-22 & 7.890$\pm$0.006 & 7.854$\pm$0.017 \\
GJ~876                          & 5.93 & 5.35 & 5.01 & 2007-12-23 & 4.806$\pm$0.016 & 4.767$\pm$0.008 \\
GJ~581                          & 6.71 & 6.10 & 5.84 & 2007-09-08 & 5.618$\pm$0.003 & 5.576$\pm$0.004 \\
GJ~849                          & 6.51 & 5.90 & 5.59 & 2007-11-25 & 5.437$\pm$0.005 & 5.427$\pm$0.006 \\
GJ~436                          & 6.90 & 6.32 & 6.07 & 2008-06-16 & 5.909$\pm$0.005 & 5.871$\pm$0.008 \\
GJ~86\tablenotemark{*}          & 4.79 & 4.25 & 4.13 & 2007-11-14 & 4.116$\pm$0.004 & 4.158$\pm$0.004 \\
HD~3651                         & 4.55 & 4.06 & 4.00 & 2008-08-21 & 3.946$\pm$0.007 & 3.968$\pm$0.004 \\
HD~69830                        & 4.95 & 4.36 & 4.17 & 2007-11-25 & 4.147$\pm$0.006 & 4.192$\pm$0.004 \\
HD~14751                        & 4.41 & 4.03 & 3.93 & 2007-09-13 & 3.904$\pm$0.006 & 3.909$\pm$0.003 \\
47~UMa                          & 3.96 & 3.74 & 3.75 & 2007-12-28 & 3.582$\pm$0.008 & 3.592$\pm$0.004 \\
55~Cnc                          & 4.77 & 4.27 & 4.02 & 2007-11-24 & 4.081$\pm$0.012 & 4.117$\pm$0.006 \\
ups And                         & 3.18 & 2.96 & 2.86 & 2008-03-09 & 2.840$\pm$0.008 & 2.858$\pm$0.005 \\
51 Peg                          & 4.66 & 4.23 & 3.91 & 2007-12-26 & 3.950$\pm$0.007 & 3.967$\pm$0.005 \\
HD~160691                       & 4.16 & 3.72 & 3.68 & 2007-09-13 & 3.579$\pm$0.005 & 3.602$\pm$0.003 \\
\enddata
\tablecomments{Stars marked with an (*) are binaries, with separations below our resolution. The photometry given is the combined photometry of both members.}
\end{deluxetable*}

\subsection{Electronic Artifact Characterization and Second PSF Subtraction}\label{elec}

After subtracting the dither-position-dependent PSF, electronic artifacts become visible in each frame. There are several effects, shown in Figure~\ref{electronicartifacts}. ``Muxstripe" is a ``jail-bar" pattern caused by excess charge in a pixel (from a bright source) unbalancing the multiplexer (readout) channel pedestals. This is superimposed with a repeating decay pattern in the lower half of the image. ``Muxbleed" is a horizontal bleeding of charge along the row(s) and readout channel(s) containing excess-charge pixels. ``Pulldown" is a central-column bias shift caused by a pixel with excess charge. We characterized these effects in sequence by creating and subtracting a correction matrix of each artifact before moving to the next: first, we masked off a circle with radius out to 10\% saturation from the center of the PSF in order to reduce the PSF subtraction residuals that get included in our calculations, then removed each artifact type in turn.

The muxstripe is a difference in pedestal value per multiplexer channel (every fourth column of the image is read by a different channel), affecting the top and bottom of the image (above and below the bright source) differently. Similarly, the muxbleed affects every fourth pixel of each of the $\sim$3 rows surrounding the core of a bright source differently. Both of these effects are characterized by the median value of the group of affected pixels, which we evaluated and then assigned to the corresponding area in the correction matrix. For example, to characterize the segment of the muxstripe in the bottom of the image in the first multiplexer channel, the median of all of the pixels comprising columns 0, 4, 8, ..., 28 from rows 0 through the row containing the central star was assigned to the correction matrix in those locations. This was done similarly for the other channels and for the muxbleed. This matrix was then subtracted from the PSF-masked image, to prepare for the next artifact to be removed.

The decay artifact does not occur in every image, nor every readout channel, but when it occurs, it again affects every fourth column of the lower half (below the bright source) of the image similarly. Previously \citep{marengo09}, we had removed this artifact by fitting an exponential or linear function to each affected column, separately. We discovered, however, that the artifacts are better removed by assuming the pattern to be the same exponential function in every affected column and subtracting the median of the pixels in each row of those columns. Affected readout channels were flagged manually and the median of each row in those columns was saved as a new correction matrix. This matrix was subtracted from the muxbleed/muxstriped-removed image.

The pulldown always occurs in the same column as the core of the target star, and is different above and below the star. When the core of the target star falls within $\sim$0.25 pixels of the boundary between pixels, the pulldown occurs on two adjacent columns. We modeled the pulldown as the median of the column(s) containing the target star, top and bottom separately, then saved it as a third correction matrix.

At this point, we added the previous three correction matrices together to obtain a complete matrix of the pixel-dependent electronic artifacts in each BCD (result shown in Figure~\ref{process}, panel c). Subtracting this matrix from the PSF-subtracted image gives us a visual check to show all artifacts have been removed (Figure~\ref{process}, panel d).

\begin{figure*}[h]
\includegraphics[width=\textwidth]{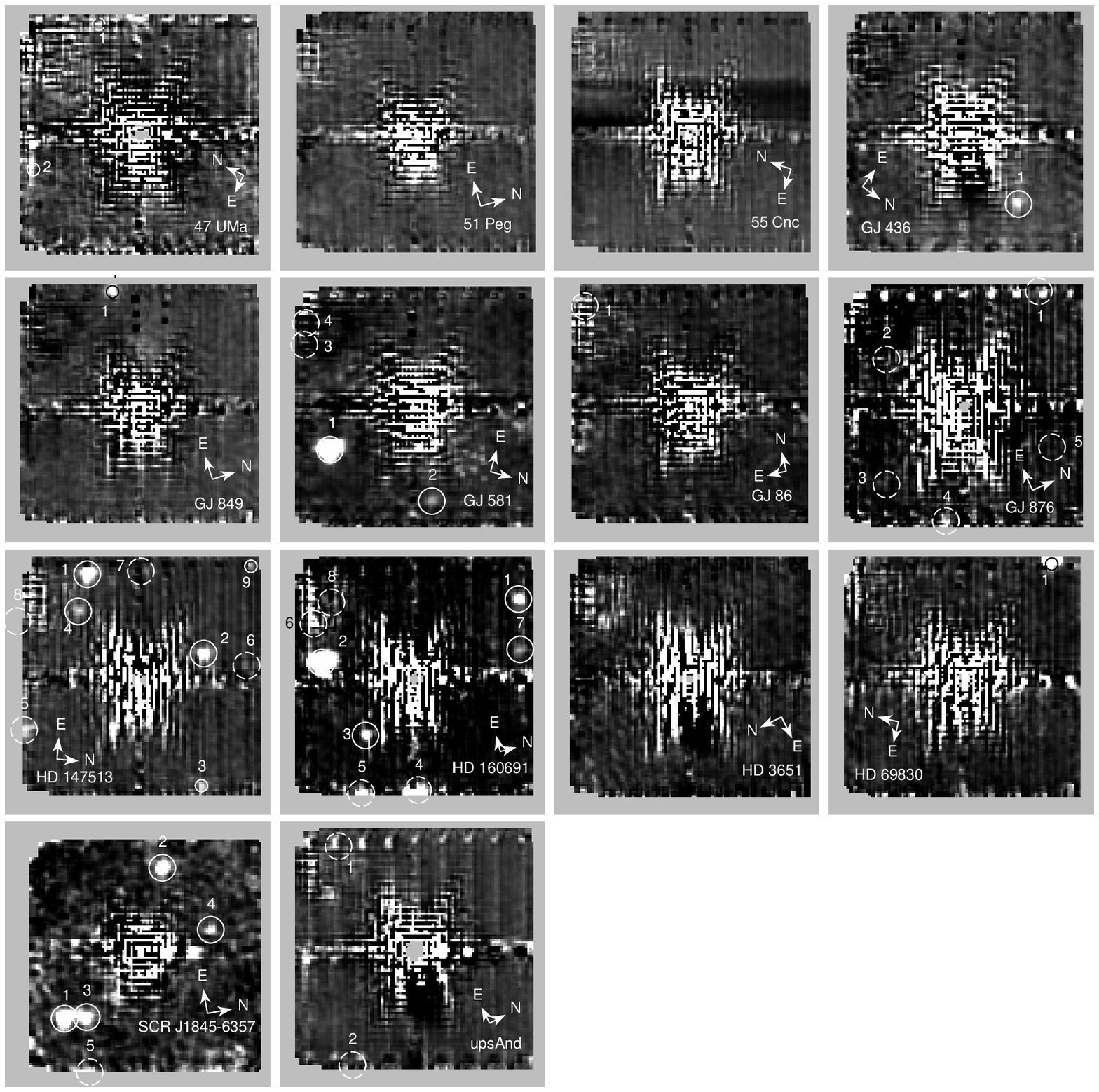}
\caption{Final 3.6~$\mu$m mosaics, with circles indicating the position of point sources detected within the subarray field of view in at least one IRAC band in either subarray or full-frame images. The size of the circle demonstrates the aperture size used for photometric measurements. Solid circles indicate detections in that frame. A  dashed circle indicates the location of a detection in different band. Unmarked bright spots are residual artifacts. Color scale is squared and adjusted for optimal contrast. The pixel scale is $0.407^{\prime\prime}$/pix. Field of view is $44^{\prime\prime}\times44^{\prime\prime}$.\label{ch1regions}}
\end{figure*}

\begin{figure*}[t]
\includegraphics[width=\textwidth]{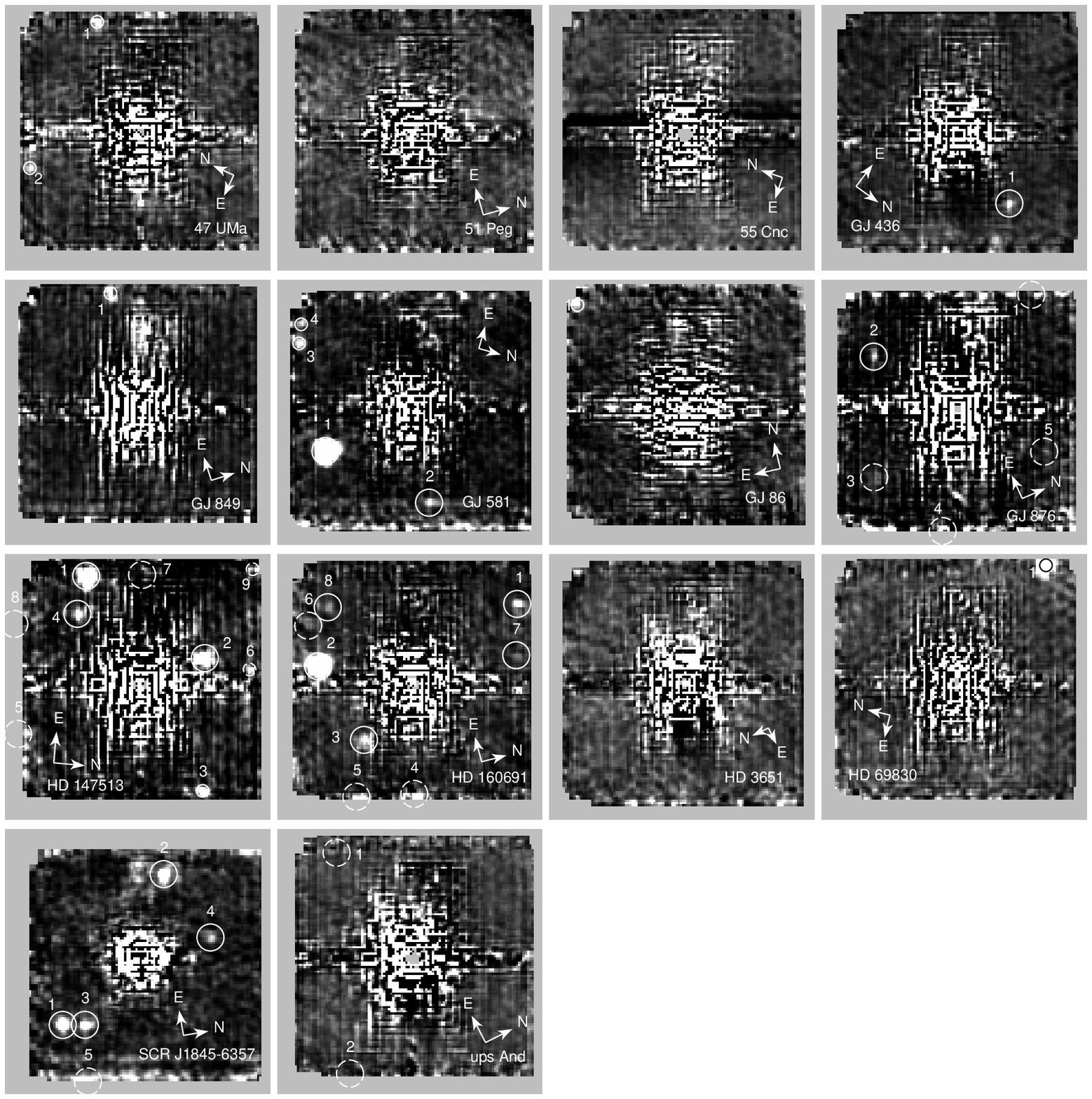}
\caption{Final 4.5~$\mu$m mosaics, with circles indicating the position of point sources detected within the subarray field of view in at least one IRAC band in either subarray or full-frame images. The size of the circle demonstrates the aperture size used for photometric measurements. Solid circles indicate detections in that frame. A  dashed circle indicates the location of a detection in different band. Unmarked bright spots are residual artifacts. The pixel scale is $0.407^{\prime\prime}$/pix. Color scale is squared and adjusted for optimal contrast. Field of view is $44^{\prime\prime}\times44^{\prime\prime}$.\label{ch2regions}}
\end{figure*}

To mitigate contamination from residual artifacts in our PSFs, we subtracted the artifact correction matrix for each image from its BCD and remade new PSFs from the result. Because all pixel dependency had been removed, and because all stars were observed with the same dither pattern, the individual dither images could then be co-added to create sub-sampled images and PSFs. We co-added the artifact-cleaned images at $10\times$ subsampling ($0.122^{\prime\prime}$/pix), then created a PSF for each star using the process described in Section~\ref{1stpsf} using the other sub-sampled images (again omitting 55~Cnc and the star itself). With nine spatially distinct dither positions, each pixel can be sub-sampled into at most nine (3$\times$3) sub-pixels, so after subtracting each star's model PSF from its final mosaic, we rebinned the images back to $3\times$ sub-sampling ($0.407^{\prime\prime}$) for analysis (Figures~\ref{ch1regions} \& \ref{ch2regions}).

Combining these images with those from PID 30666, we created and released a $0.24^{\prime\prime}$/pix subarray PSF, available at the Spitzer Science Center website.

\subsection{Source detection and photometry}\label{sourcephotometry}

Once the PSF and electronic artifacts are removed and the images are co-added, other sources become readily visible (see Figures~\ref{ch1regions} \& \ref{ch2regions}). Given the small number of sources in each image, we visually inspected each and measured the photometry of every area that seemed brighter than the background, using a custom procedure based on DAOPHOT. We used an aperture radius of two IRAC pixels (2.44$^{\prime\prime}$) where possible, or one (1.22$^{\prime\prime}$) for sources near the edge of the final images. Sky annuli were chosen to fit the background while avoiding excessive PSF subtraction noise for the typical source, and had an inner radius of 3.17$^{\prime\prime}$ (1.59$^{\prime\prime}$) with an outer radius of 5.12$^{\prime\prime}$ (2.56$^{\prime\prime}$) for the larger (smaller) apertures. Aperture corrections were calculated as the difference in magnitudes between the flux inside each aperture listed above versus the flux inside a 10-IRAC-pixel aperture placed on a Vega-scaled PSF. These corrections in 4.5~$\mu$m are $-0.24$ mag for the larger aperture, $-0.94$ mag for the smaller. Any source calculated to have a local signal-to-noise ratio less than 3 was rejected. All sources detected in 3.6~$\mu$m and 4.5~$\mu$m are shown Figures~\ref{ch1regions} and \ref{ch2regions}, respectively. Table~\ref{sourcephot} shows the aperture-corrected photometry of all detected sources in the frame of each target star. For sources detected in only one band, the limiting magnitude at that location in the other band is given as a photometric upper limit. Four objects (47~UMa-1, HD~160691-8, GJ~86-1, and GJ~581-3) have potential colors that can be as red as a T5 or later brown dwarf, due to non-detections in 3.6~$\mu$m.

The subarray field of view is too small for the analysis pipeline to perform automatic pointing refinement. No subarray frames had sufficient 2MASS sources to perform manual astrometric calibration, and our observations were taken before the ``peak up" function was available. Therefore, the accuracy by which the WCS coordinates are known is limited by Spitzer's star tracker, plus pointing drift, which is especially important for short exposures like subarray mode, and jitter; the combination of these is on the order of 1$^{\prime\prime}$. We performed relative astrometry on each source by computing its centroid and calculating the distance and position angle relative to the centroid of the central source (Table~\ref{sourcephot}). Because the PSF is pixelated and has internal structure, the centroid can converge in a slightly different location depending on the subsampling of the PSF due to the source falling on different parts of a pixel. To calculate the error of our relative astrometry, we randomly shifted our 0.24$^{\prime\prime}$/pix model PSF 1000 times, each time rebinning to 0.407$^{\prime\prime}$/pix and comparing the known shift to the difference in measured centroids. The average of the resulting distribution is 0.05 pixels, or 0.02$^{\prime\prime}$. This represents the best-case radial distance error for high signal-to-noise, isolated sources. For more crowded, dim sources, the convergence point of the centroid also depends on the initial guess. To calculate this error, we randomly picked 1000 starting locations within a radius of 2.1 pixels (half the FWHM of IRAC's band 2 at 0.407$^{\prime\prime}$/pix) around each source, and calculated the centroid using a box size of 6 pix. This produced a distribution of between 1 and 6 ``centroids." We report the center of the distribution as the source position and the RMS spread as the initial-guess error. This error was on the order of 0.1$^{\prime\prime}$ and was added in quadrature with the PSF-discretization error to find our total relative astrometry error, reported per source in Table~\ref{sourcephot}.

\subsection{Full-frame Analysis}\label{fullframe}

To supplement and confirm our subarray results, we performed follow-up photometry for all stars that had full-frame images available in the Spitzer Heritage Archive (listed in ``full" columns of Table~\ref{sourcephot}). These observations were performed with 30-second frame times and 5 Gaussian dithers. We used IRACproc to create mosaics and subtract the target stars' PSFs, then we performed aperture photometry. To isolate the typical source from artifacts and other sources of noise, we used a 2.83$^{\prime\prime}$ aperture radius, with a sky annulus from 2.83$^{\prime\prime}$ to 4.72$^{\prime\prime}$ on all sources visible within the subarray-equivalent field of view. We restricted our analysis to this region because everything outside the subarray field of view in the full frame images had already been discounted by a previous proper motion search for widely separated companions without PSF subtraction \citep{patten05}.

\begin{figure*}[t]
\includegraphics[width=\textwidth]{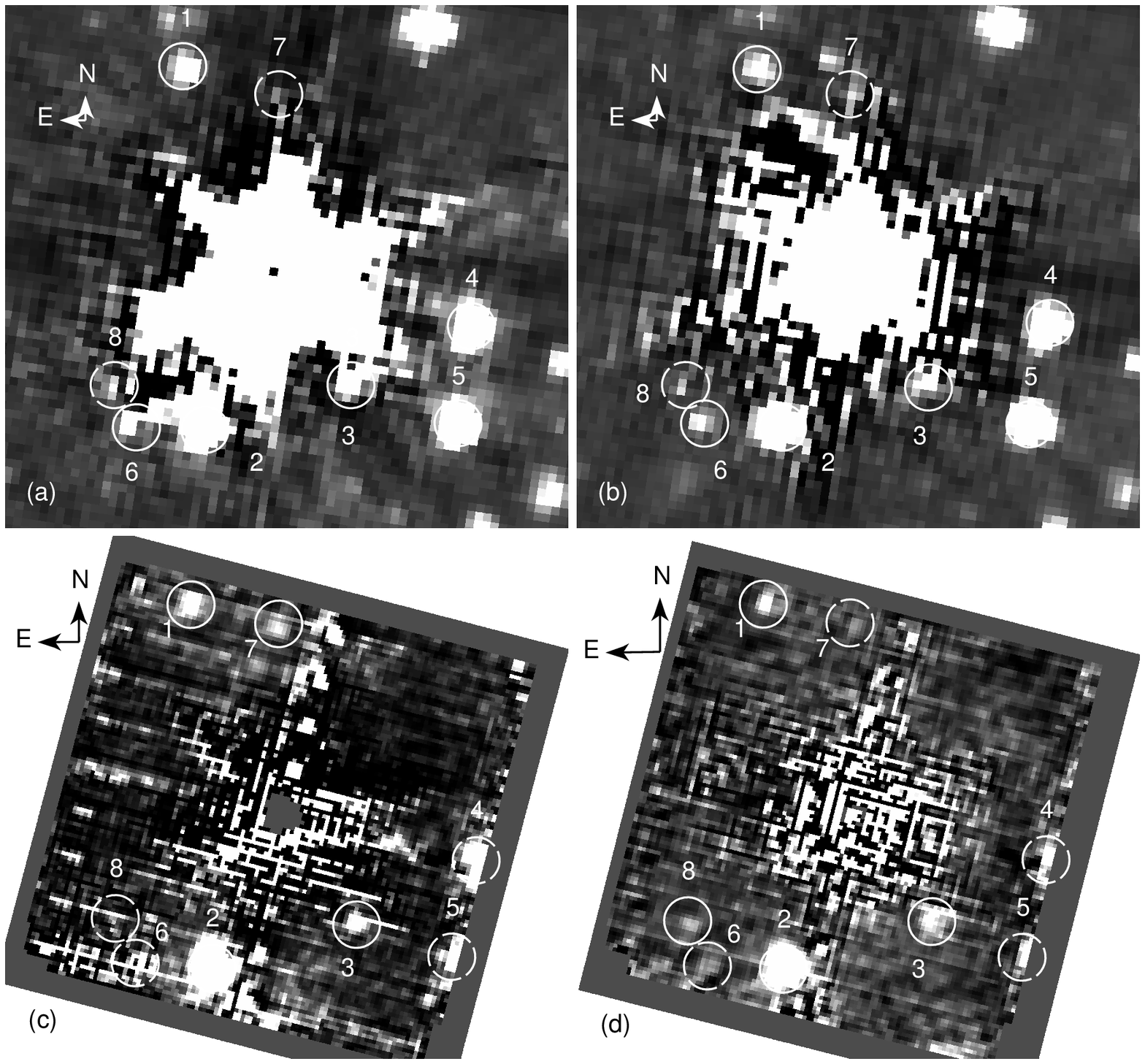}
\caption{Comparison of HD~160691 in full frame 3.6~$\mu$m (a), 4.5~$\mu$m (b) and subarray 3.6~$\mu$m (c), 4.5~$\mu$m (d) fields. Solid circles indicate detections in that frame. A  dashed circle indicates the location of a detection in different band. Sources 2, 3, and 8 are detected in subarray but overwhelmed by residuals in full frame. Sources 4 and 5 were originally thought to be noise in subarray, but are point sources in full frame. Sources 6 and 7 are at the sensitivity limit in subarray and affected by PSF residuals in full frame. Unmarked sources fall outside the subarray field of view and were analyzed previously.\label{hd160691}}
\end{figure*}

Several sources detected in subarray images were found to have counterparts in the full-frame images, including sources near the edge of the subarray frame that we originally thought to be noise (see sources at the edges of HD~160691, HD~147513, and SCR~J1845-6357). Some sources appearing in full-frame images were below the sensitivity of the subarray images. As expected, sources detected near the target star in subarray were overwhelmed by PSF residual noise in the full-frame images, notably HD~147513-2, SCR~J1845-6357-2, and GJ~876-7. Figure~\ref{hd160691} shows a comparison of objects seen in the subarray field of view between subarray and full-frame images for one parent star.

\begin{figure*}[t]
\includegraphics[angle=-90,width=\textwidth]{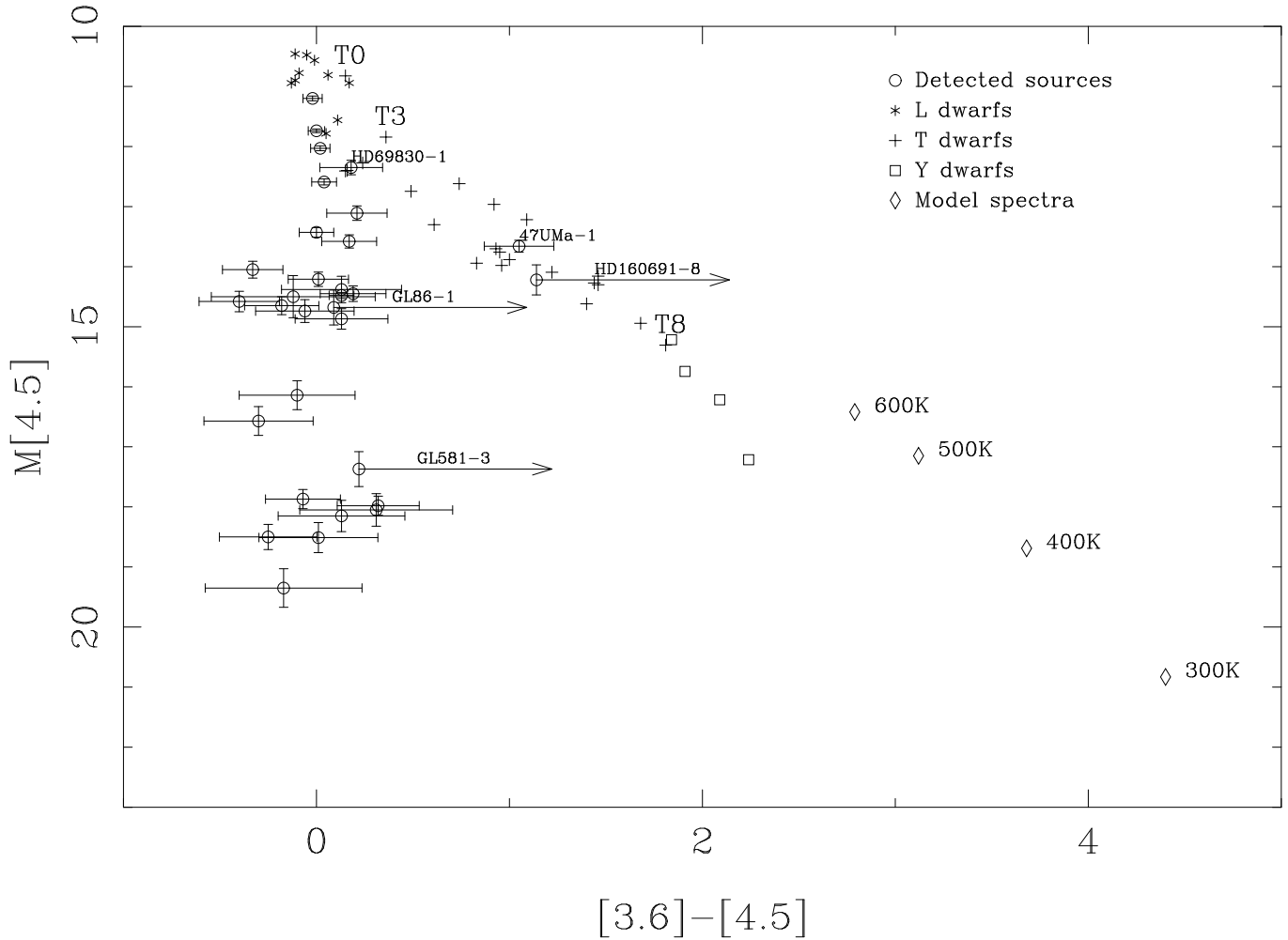}
\caption{Mid-infrared HR diagram of all detected sources (circles) within the subarray fields of view, assuming each to be at the distance of its target star. The lowest-error photometry from either subarray or full frame in each band is displayed (circles). Temperatures and absolute magnitudes from non-equilibrium substellar model spectra \citep{hubeny07} are plotted with diamonds; asterisks show a selection of published L-dwarfs, while published T- and Y- dwarfs are displayed as plus signs and squares, respectively (L \& T photometry from \citealt{patten06}; Y photometry from \citealt{ashby09}).\label{hrdiagram}}
\end{figure*}

Taking the lowest-error photometry in each band from either subarray or full frame, and assuming all of these sources to be at the same distance as their parent star, we plotted their [3.6]$-$[4.5] color versus 4.5 $\mu$m absolute magnitude (see Figure~\ref{hrdiagram}) to compare with model substellar-mass objects of various temperatures \citep{burrows02} and previously detected L and T dwarfs \citep{patten06}.

\section{Sensitivity}\label{sensitivitysection}

\begin{figure*}[t]
\includegraphics[width=\textwidth]{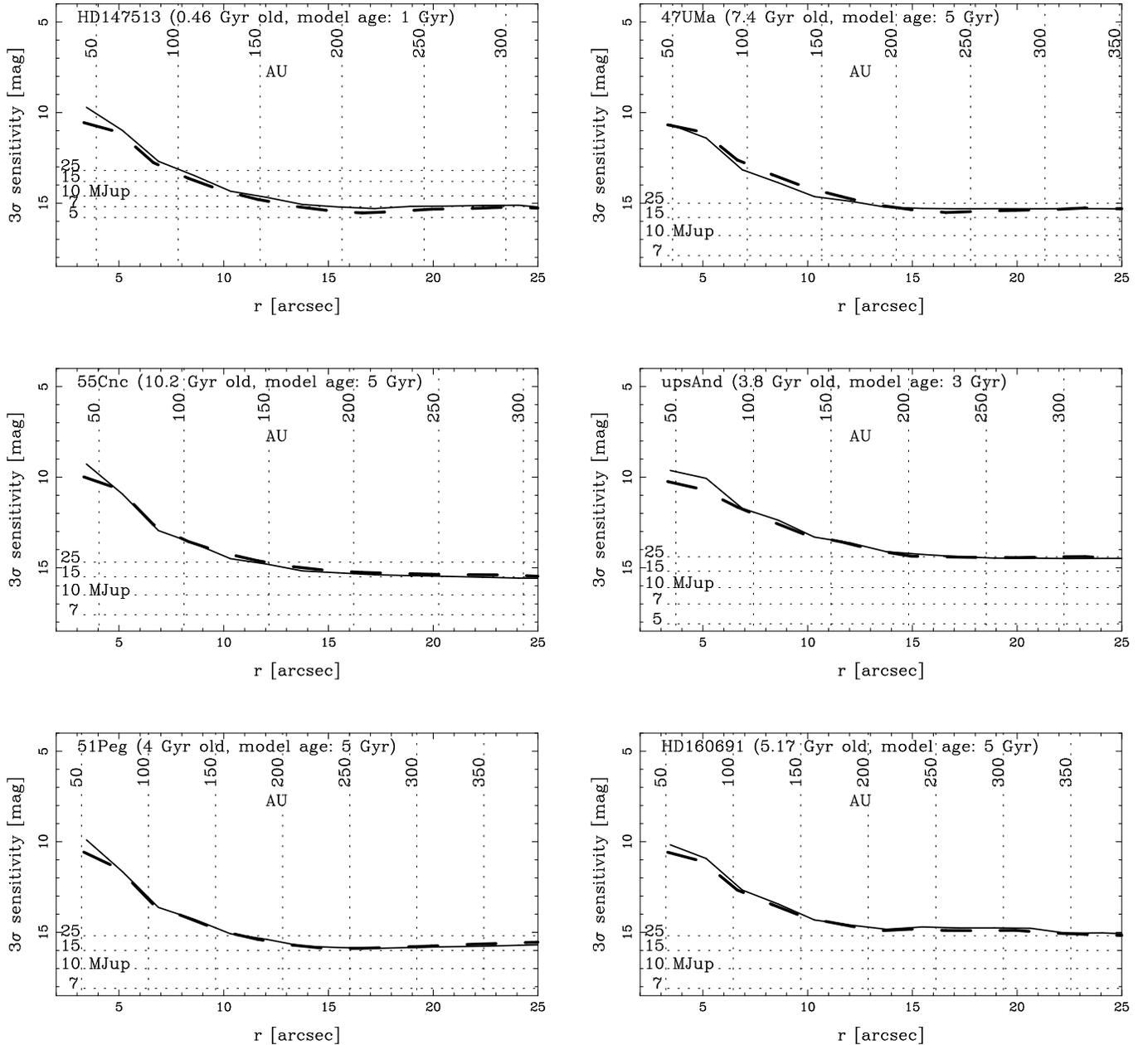}
\caption{3.6~$\mu$m (dashed) and 4.5~$\mu$m (solid) radial sensitivity curves are shown for each star. Vertical dotted lines show projected semi-major axis related to radial separation for each star. Horizontal dotted lines show model 4.5~$\mu$m magnitude estimates from \citet{burrows03} for a range of potential companion masses, given the estimated age of each star.
\label{sensitivity}}
\end{figure*}

\begin{figure*}[t]
\includegraphics[width=\textwidth]{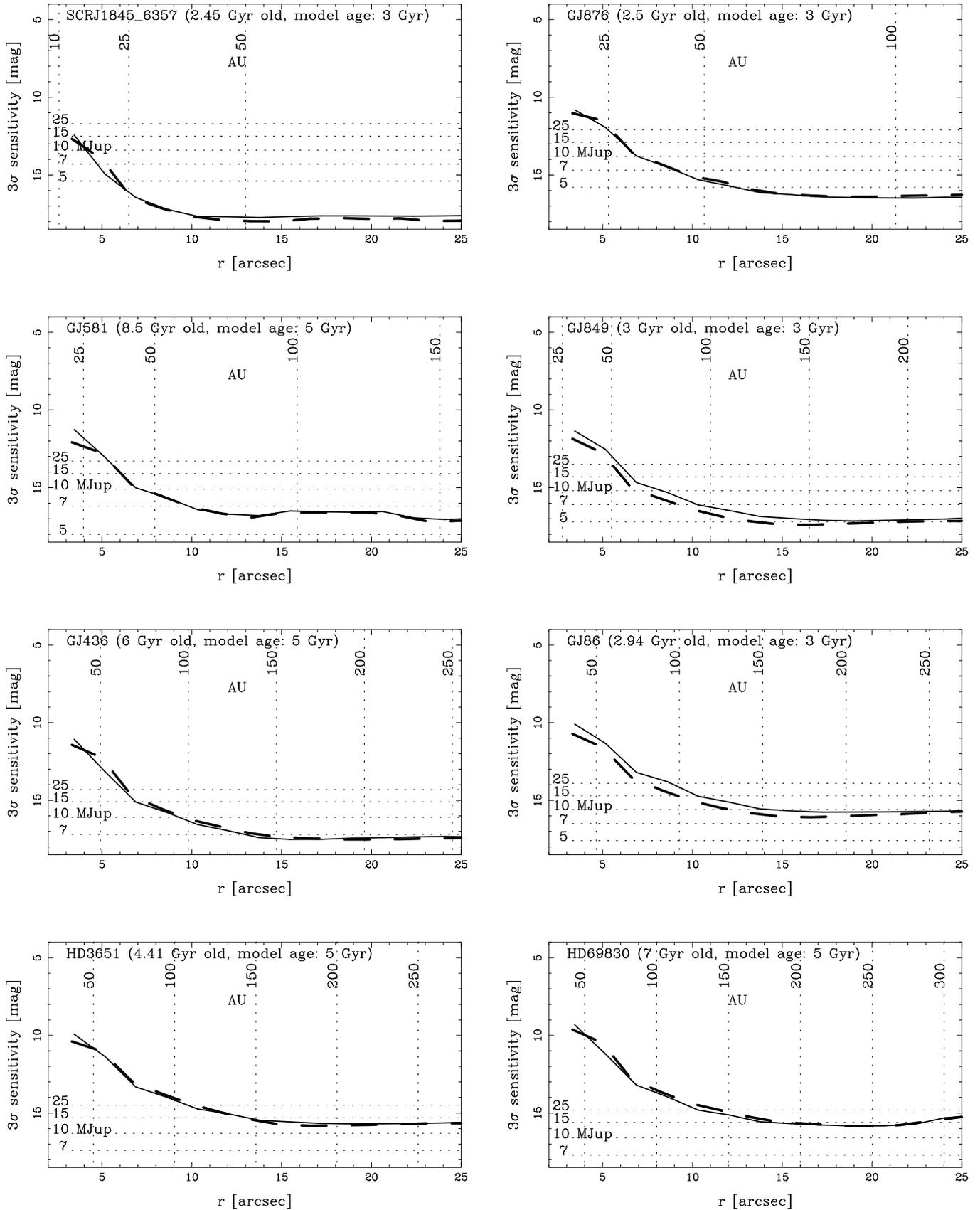}
\caption{Figure~\ref{sensitivity}, continued.\label{sensitivity2}}
\end{figure*}

We calculated sensitivity limits for each of our targets as a function of projected distance from the central star. Our photometric sensitivity at any location within an image is limited by the local noise level around that point. The noise level within the frame is generally highest at the location of the target star and decreases radially outward, but with azimuthal deviations due to PSF spikes and other features. We generated azimuthally averaged sensitivity curves, giving the Noise-Equivalent Flux Density (NEFD) for given radial distances from the target star (Figures~\ref{sensitivity} and \ref{sensitivity2}). We calculated the NEFD in each point of the PSF-subtracted image as the total RMS noise flux over an aperture with diameter equal to the FWHM of the stellar PSF (1.66$^{\prime\prime}$ for 3.6~$\mu$m, 1.72$^{\prime\prime}$ for 4.5~$\mu$m). Overlaid on these plots are the estimated magnitudes of planets with different masses, derived from non-irradiated extrasolar planet models with a given age from \citet{burrows02}, scaled to the distance of the primary star. The only models available from \citet{burrows02} for ages comparable to our objects are 1, 3, and 5~Gyr, so we chose the model age closest to the age of each star. For ages greater than 5~Gyr, we used the 5~Gyr model, though this underestimates the mass of potential companions. The age we used for the model is listed with the star's estimated true age above each plot.

Also available were the BT-SETTL models by \citet{allard11}, which have a greater range of available ages, but much higher lower limit of $\sim50$~M$_{J}$ for all but the youngest isochrones, making them insufficient to match our sensitivity. We compared the Burrows and Allard models for the same object age and found they give comparable magnitudes for a given object mass. For the smallest BT-SETTL object available for the 10~Gyr isochrone, 63~M$_{J}$, the difference in 4.5~$\mu$m model magnitude from 5 to 10~Gyr is 0.42. Less-massive objects should have lost more of their primordial heat by the time they reach 5~Gyr old, thus they would be expected to dim by less than this amount over the following 5~Gyr, so we feel confident that we are not significantly overestimating our mass sensitivity for objects older than 5~Gyr.

Our best sensitivity occurs for nearby, young systems like SCR~J1845-6357, where we detected no $>$5~M$_{J}$ objects at projected separations between 25-80 au from its parent, and GJ~876, for which our limit is of a $>$6~M$_{J}$ object at projected separations from 50-100 au. For the older, more distant stars like $\upsilon$~And, 55~Cnc, and 51~Peg, we detected no objects greater than 25~M$_{J}$ between projected separations of roughly 200-300 au, with the caution that our mass model sensitivity is overestimated for systems older than 5~Gyr. Our lower-limit sensitivity around the remaining stars varies from 7 to 15~M$_{J}$ in the range of projected separations of 50 to 250 au.

\section{Color-Selected Companion Candidates}\label{detectedsources}

\subsection{47 UMa-1}\label{47uma}

47~UMa has three previously confirmed RV planets of roughly Jupiter mass \citep{gregory10}. The outermost, 47~UMa~d, has the highest eccentricity of the three (e=0.16$^{+0.09}_{-0.16}$). This suggested eccentricity level is difficult to explain without an external perturber. Our subarray observations revealed 47~UMa-1, a source that appears in 4.5~$\mu$m only, and has upper limit colors ([3.6]-[4.5]$>1$ and M[4.5]=14.7) comparable to a brown dwarf at the distance of the primary. If the object is a bound companion, it would be located at a projected separation of $\sim$300 au (22$^{\prime\prime}$) from 47~UMa. If it is not bound, its color indicates it may be a background red giant or galaxy (\citealt{reiter14}; \citealt{stern07}).

We found archival full frame observations of 47~UMa, performed Apr 20, 2004 as part of PID 347513231. 47~UMa-1 was detected in the full-frame images but is unfortunately directly on top of a strong pulldown artifact in both bands. Full frame photometry confirms this object to be more than a magnitude red in [3.6]-[4.5]. Assuming it to be at the same distance as 47~UMa gives it [3.6]-[4.5] vs. M[4.5] coordinates matching a T5 dwarf, with a model mass of $\sim25~M_{J}$ (see Figure~\ref{hrdiagram}).

\begin{figure}[h]
\includegraphics[angle=-90,width=\columnwidth]{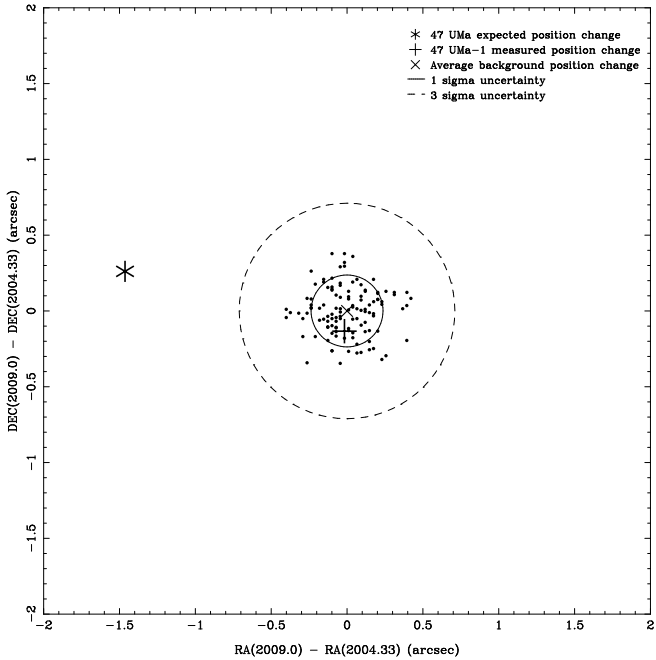}
\caption{The measured (plus sign) change in position of 47~UMa-1 from 2004.33 to 2009.0, compared with the average change in position ($\times$) of field sources (points) and the expected change in position of 47~UMa (asterisk). We subtracted the average change in position from all measured points to correct for any systematic offset between epochs. It is unlikely that 47~UMa-1 is a co-moving companion of 47~UMa. \label{propermotion}}
\end{figure}

Another full-frame observation of 47~UMa (4.5~$\mu$m only) was made Dec 23, 2008 (PID 332319740), so we attempted to verify common proper motion. We measured the proper motion of 118 sources surrounding 47~UMa to compare to the motion of source~1 and the expected position change of the primary over the 4.67 years between the two observations (Figure~\ref{propermotion}). The measured position change of 47~UMa-1 is more than 3$\sigma$ from that expected of 47~UMa, making it unlikely to be a co-moving companion.

The other source in the same frame, 47~UMa-2, was affected too much by artifacts in both full frame archival epochs to obtain reliable photometry or proper motion. Undetected objects at projected separations between 200-350 au are constrained to have a model dependent mass less than 15~M$_{J}$, with the caveat that we used the 5 Gyr model for this 7.4-Gyr-old star.

\subsection{HD~160691-8}

The four known RV planets of the HD~160691 system all have e$\approx$0.1 \citep{pepe07}. \citet{koriski11} noted that this system has a 2/1 period commensurability, which could indicate mean motion resonance, which in turn could explain the eccentricities of the system without invoking an external body. We found one source, HD~160691-8, with only an upper limit color, potentially similar to a $\sim$T7 dwarf with mass near 25~M$_{J}$. If a bound companion, it would be located at a projected separation of $\sim$300 au (21$^{\prime\prime}$) from HD~160691. In the full-frame images taken Sept 9, 2004 (PID 240828495), this source lies directly on top of PSF spike residuals (see Figure~\ref{hd160691}), so we were not able to calculate its full-frame photometry.

We found a model-dependent upper mass limit of 25~M$_{J}$ for unseen objects at a projected distance between 200 and 400 au from the primary.

\subsection{GJ~581-3}

Of the four known RV planets around GJ~581, the outermost two, GJ~581d and GJ~581e, have e=0.205 and e=0.32, respectively \citep{forveille11}. \citet{baluev12} claims that the existence of planet d is questionable but confirms b, c, and e. Planet e is the least massive of the four (respectively, they are 0.05~M$_{J}$, 0.017~M$_{J}$, 0.019~M$_{J}$, and 0.0061~M$_{J}$), so it may be likely that some earlier scattering event led to its current orbit.

If GJ~581-3 is a bound object, its 4.5~$\mu$m flux implies it would be a Y~dwarf, with model mass of 7-10~M$_{J}$, located at a projected separation of $\sim$150 au (23$^{\prime\prime}$) from GJ~581. We found no objects with mass $>7~M_{J}$ between projected separations of 75-150 au from the primary. Our mass limits for this star are underestimates: our oldest mass model was 5 Gyr and this star has an age of 8.5 Gyr. GJ~581 had no available archival full-frame observations.

\subsection{GJ~86-1}

GJ~86 has one marginally detected source in the very corner (26$^{\prime\prime}$ from center) of the 4.5~$\mu$m subarray frame. This source is not detected at 3.6~$\mu$m, due to strong PSF-subtraction residuals caused by the IRAC filter ``ghost." Nevertheless, the limiting sensitivity at that location ([3.6]$>$14.94) implies a color [3.6]-[4.5]$>$0.9. If the source is a bound companion at the distance of the primary, its 4.5~$\mu$m magnitude and color limit would be consistent with a ~T7 dwarf. However, the high level of noise, even in the 4.5~$\mu$m frame, suggests the source to be a spurious detection. We attempted to measure full-frame photometry of GJ~86-1, but both its WCS-coordinate location and where it would be if it shared common proper motion with GJ~86 are overwhelmed by the pulldown in the full frame.

\subsection{HD~69830-1}

The only source detected near HD~69830 is cut off at the edge of the frame. Its estimated photometry puts it at the [3.6]-[4.5] color boundary between stars and T-dwarfs (Figure~\ref{hrdiagram}). There were no full-frame archival observations of this object in order to obtain better IRAC photometry, but it does have 2MASS photometry available. HD~69830 has visibly moved between the 2MASS epoch and our observations taken in Nov 2007, but HD~69830-1 has not, so they do not share common proper motion.

\citet{tanner10} reported a candidate M companion around HD~69830, with relative coordinates (pointing angle, separation) of -152.5$\degr{}$, 9.99$^{\prime\prime}$ and K=15.46. Assuming its K magnitude to be similar to [3.6], this object is just below our sensitivity at that position (see Figure~\ref{sensitivity}).

\section{Discussion and Summary}\label{discussion}

Using the combined photometry from our subarray images and the available archival full-frame images, we found four potentially interesting candidates: one with matching brown dwarf colors and three objects with compatible [3.6]$-$[4.5] color lower limits. In absence of common proper motion confirmation, we are unable to determine if these candidates are true companions, or unrelated background quasars or mass-losing giants (see \citealt{stern07}; \citealt{reiter14} for typical colors of background red sources). This ambiguity could, in principle, be resolved by acquiring JHK photometry (see \citealt{marengosanchez09}), albeit new near-IR data capable of detecting these objects would also likely provide accurate astrometry, sufficient to test for common proper motion.

We presented details of our PSF- and artifact-subtraction procedure, in which we were able to automatically co-align images within 0.0027$^{\prime\prime}$ for purposes of PSF subtraction and effective removal of electronic artifacts. As anticipated, subarray observations allowed a smaller inner working angle than full-frame observations of the same field of view, which were dominated by PSF subtraction residuals.

For all stars, we calculated upper limits on unseen T- or Y-dwarfs within the mass and projected semi-major axis ranges given in Figures~\ref{sensitivity} and \ref{sensitivity2}. Outside of 10$^{\prime\prime}$ from most sources, our sensitivity is in the range of 10~M$_{J}$. Our best sensitivity is for close, young stars like GJ~849 (5~M$_{J}$) and SCRJ~1849-6357 ($<5~M_{J}$). Our worst sensitivity (25~M$_{J}$) was for $\upsilon$~And due to its brightness: artifacts overwhelmed the image.

We could not confirm the existence of any new wide planetary or brown dwarf companions around the stars in our sample. We have ruled out a section of parameter space, but there is a degeneracy for perturbers between mass and semi-major axis. It is still possible for less-massive companions closer than our inner working angle to exist (ongoing systematic searches with new ground-based adaptive optics systems will explore the remaining parameter space), or there could be some other reason for the measured high eccentricities of planets in those systems.

Follow up observations are necessary for HD~160691-8, GJ~86-1, and GJ~581-3 to obtain 3.6~$\mu$m photometry and check for common proper motion. If the candidates are confirmed, numerical simulations will be necessary to determine whether they can account for the high eccentricities of planets in their respective systems.

\acknowledgments
Thanks to Brett Kail for helping to brainstorm ideas on image co-alignment. This work is based on observations made with the Spitzer Space Telescope, which is operated by the Jet Propulsion Laboratory, California Institute of Technology, under a contract with NASA. Support for this work was provided by NASA through an award issued by JPL/Caltech. A.H., M.M., and J.C. were partially supported by the U.S. National Science Foundation under Award No. 1009203. J.C. received support from the South Carolina Space Grant Consortium and the Research Corporation for Science Advancement (Award No. 21026).

{\it Facilities:} \facility{Spitzer}.

\clearpage
\begin{landscape}

\begin{deluxetable}{lcccccccccccc}
\tablefontsize{\footnotesize}
\tablecaption{Photometry of detected sources\label{sourcephot}}
\tablewidth{0pt}
\tablehead{
\colhead{Parent star} & \colhead{Ref num} & \colhead{RA} & \colhead{Dec} & \colhead{Rad}  & \colhead{PA} & \colhead{J} & \colhead{H} & \colhead{K} & \colhead{Sub[3.6]} & \colhead{Sub[4.5]} & \colhead{Full[3.6]} & \colhead{Full[4.5]}
}
\startdata
GJ~876           &  1   &   22:53:18.216 & -14:15:32.80  &  26.5$\pm$0.2$^{\prime\prime}$ & 122.2$\pm$0.1$\degr{}$ &   --   &	--	    &	--	    &	--            &  --			   & 17.08$\pm$0.27 & $>19.37$       \\
GJ~876           &  2   &   22:53:18.254 & -14:16:04.80  &  17.5$\pm$0.3$^{\prime\prime}$ & 212.9$\pm$1.3$\degr{}$ &   --   &    --     &   --      &  --             & 16.60$\pm$0.27 & 16.64$\pm$0.20 & 16.50$\pm$0.26 \\
GJ~876           &  3   &   22:53:16.760 & -14:16:14.49  &  20.7$\pm$0.4$^{\prime\prime}$ & 291.8$\pm$0.6$\degr{}$ &   --   &      --   &    --     &  --             & --             & 17.25$\pm$0.29 & 17.09$\pm$0.37 \\
GJ~876           &  4   &   22:53:16.049 & -14:16:07.01  &  21.8$\pm$0.4$^{\prime\prime}$ & 327.4$\pm$0.3$\degr{}$ &   --   &   --      &    --     &   --            & --             & 16.71$\pm$0.29 & 16.41$\pm$0.27 \\
GJ~876           &  5   &   22:53:16.296 & -14:15:45.50  &  17.5$\pm$0.3$^{\prime\prime}$ &  34.4$\pm$0.3$\degr{}$ &   --   &   --      &   --      &  $>18.38$       & $>17.83$       & 14.92$\pm$0.13 & 17.01$\pm$0.32 \\
HD~147513        &  1   &   16:24:02.914 & -39:11:46.43  &  26.5$\pm$0.2$^{\prime\prime}$ & 205.2$\pm$0.6$\degr{}$ &  13.06	&	12.84	&	13.20   &  13.07$\pm$0.10 & 13.19$\pm$0.13 & 13.16$\pm$0.04 & 13.12$\pm$0.05 \\
HD~147513        &  2   &   16:24:01.805 & -39:11:23.93  &  13.4$\pm$0.2$^{\prime\prime}$ & 123.4$\pm$0.2$\degr{}$ &  --    &	--      &	--      &  13.85$\pm$0.10 & 13.64$\pm$0.12 &    --          & --             \\
HD~147513        &  3   &   16:23:59.686 & -39:11:22.02  &  27.7$\pm$0.1$^{\prime\prime}$ &  28.2$\pm$0.3$\degr{}$ &  13.10	&	13.18	&	14.31	&  14.76$\pm$0.21 & 15.04$\pm$0.32 & 14.75$\pm$0.10 & 14.74$\pm$0.12 \\
HD~147513        &  4   &   16:24:02.338 & -39:11:47.15  &  19.5$\pm$0.1$^{\prime\prime}$ & 218.0$\pm$0.4$\degr{}$ &	--  &	--	    &	--    	&  15.05$\pm$0.22 & 14.91$\pm$0.22 & --             & --             \\
HD~147513        &  5   &   16:24:00.295 & -39:11:55.39  & 	24.4$\pm$0.1$^{\prime\prime}$ & 306.7$\pm$0.4$\degr{}$ &	--  &	--    	&	--	    &	--            &             -- & 13.96$\pm$0.04 & 13.97$\pm$0.08 \\
HD~147513        &  6   &   16:24:01.641 & -39:11:16.38  &  19.4$\pm$0.4$^{\prime\prime}$ & 104.7$\pm$0.8$\degr{}$ &	--  &	--    	&	--    	&   $>15.59$      & $>15.19$       & 15.21$\pm$0.17 & 15.27$\pm$0.19 \\
HD~147513        &  7   &   16:24:03.070 & -39:11:36.31  &  26.4$\pm$0.2$^{\prime\prime}$ & 182.6$\pm$0.5$\degr{}$ &	--	&	--	    &	--	    &   --            & --             & 15.53$\pm$0.17 & 15.40$\pm$0.17 \\
HD~147513        &  8   &   16:24:02.031 & -39:11:59.26  &  26.4$\pm$0.4$^{\prime\prime}$ & 245.9$\pm$1.1$\degr{}$ &  13.57 &	15.18   &	14.80   &   --            & --             & 15.01$\pm$0.12 & 15.18$\pm$0.15 \\
HD~147513        &  9   &   16:24:03.240 & -39:11:16.37  &  34.5$\pm$0.2$^{\prime\prime}$ & 147.0$\pm$0.1$\degr{}$ &	--	&	--    	&	--	    &  15.40$\pm$0.28 & $>15.52$       & 15.17$\pm$0.11 & 14.99$\pm$0.13 \\
47~UMa           &  1   &   10:59:26.441 & +40:26:05.10  &  25.5$\pm$0.1$^{\prime\prime}$ &  37.6$\pm$0.4$\degr{}$ &	--  &	--	    &	--	    &	$>15.54$      & 14.69$\pm$0.25 & 15.45$\pm$0.15 & 14.40$\pm$0.10 \\
47~UMa           &  2   &   10:59:28.980 & +40:26:04.88  &  23.6$\pm$0.1$^{\prime\prime}$ & 139.5$\pm$0.1$\degr{}$ &     -- &   --      &     --    &  15.12$\pm$0.24 & 15.24$\pm$0.35 &   $>18.25$     & 15.92$\pm$0.34 \\
HD~160691        &  1   &   17:44:09.617 & -51:49:42.35  &  27.1$\pm$0.2$^{\prime\prime}$ & 123.8$\pm$0.2$\degr{}$ &  12.83	&	13.07	&	14.54   &  14.59$\pm$0.15 & 14.49$\pm$0.23 & 14.70$\pm$0.09 & 14.53$\pm$0.11 \\
HD~160691        &  2   &   17:44:09.403 & -51:50:20.54  &  19.7$\pm$0.1$^{\prime\prime}$ & 232.9$\pm$0.4$\degr{}$ &  --    &	--  	&	--      &  12.03$\pm$0.04 & 12.15$\pm$0.06 & 12.13$\pm$0.03 & 12.15$\pm$0.04 \\
HD~160691   	 &  3   &   17:44:07.754 & -51:50:16.37  &  17.3$\pm$0.3$^{\prime\prime}$ & 318.1$\pm$0.1$\degr{}$ &	--	&	--	    &	--	    &  14.54$\pm$0.15 & 14.81$\pm$0.24 & 14.66$\pm$0.07 & 15.05$\pm$0.14 \\
HD~160691        &  4   &   17:44:06.403 & -51:50:09.53  &  33.4$\pm$0.1$^{\prime\prime}$ & 352.0$\pm$0.1$\degr{}$ &	--  &	--	    &	--	    &   --            & --             & 12.99$\pm$0.03 & 13.03$\pm$0.04 \\
HD~160691        &  5   &   17:44:06.667 & -51:50:19.21  &  32.5$\pm$0.1$^{\prime\prime}$ & 333.8$\pm$0.1$\degr{}$ &  12.63 &	12.43   &	12.77   &   --            & --             & 12.68$\pm$0.03 & 12.74$\pm$0.03 \\
HD~160691        &  6   &   17:44:10.234 & -51:50:20.17  &  28.8$\pm$0.4$^{\prime\prime}$ & 212.2$\pm$1.1$\degr{}$ &	--  &	--	    &	--	    &  --             & --             & 15.13$\pm$0.12 & 15.53$\pm$0.17 \\
HD~160691        &  7   &   17:44:08.602 & -51:49:44.83  &  20.0$\pm$0.3$^{\prime\prime}$ &  89.5$\pm$0.8$\degr{}$ &	--  &	--	    &	--	    &  15.58$\pm$0.28 & $>15.79$       & --             & --             \\
HD~160691        &  8   &   17:44:10.481 & -51:50:15.79  &  30.1$\pm$0.2$^{\prime\prime}$ & 201.3$\pm$0.5$\degr{}$ &	--  &	--	    &	--	    &   $>16.31$      & 15.17$\pm$0.25 & --             & --             \\
SCR~J1845-6357	 &  1   &   18:45:06.809 & -63:57:59.44  &  27.4$\pm$0.2$^{\prime\prime}$ & 324.7$\pm$0.1$\degr{}$ &	--	&	--	    &	--	    &  15.73$\pm$0.11 & 15.80$\pm$0.16 & 15.88$\pm$0.14 & 15.88$\pm$0.19 \\
SCR~J1845-6357   &  2   &   18:45:10.488 & -63:57:36.86  &  33.6$\pm$0.2$^{\prime\prime}$ & 168.4$\pm$0.2$\degr{}$ &	--	&	--	    &	--	    &  16.23$\pm$0.14 & 15.90$\pm$0.16 & 16.51$\pm$0.18 & 16.26$\pm$0.22 \\
SCR~J1845-6357	 &  3   &   18:45:06.698 & -63:57:55.51  &  26.8$\pm$0.2$^{\prime\prime}$ & 333.6$\pm$0.2$\degr{}$ &	--	&	--	    &	--	    &  16.18$\pm$0.14 & 16.43$\pm$0.21 & 16.26$\pm$0.16 & 16.18$\pm$0.24 \\
SCR~J1845-6357   &  4   &   18:45:08.546 & -63:57:30.38  &  13.7$\pm$0.2$^{\prime\prime}$ & 105.7$\pm$0.5$\degr{}$ &	--	&	--   	&	--	    &  17.48$\pm$0.27 & 17.28$\pm$0.32 & 17.10$\pm$0.25 & 17.03$\pm$0.35 \\
SCR~J1845-6357	 &  5   &   18:45:05.128 & -63:57:56.62  &  48.7$\pm$0.4$^{\prime\prime}$ & 344.4$\pm$0.6$\degr{}$ &	--  &	--	    &	--	    &   --            &   --           & 16.45$\pm$0.18 & 16.44$\pm$0.25 \\
GJ~86            &  1   &   02:10:29.273 & -50:48:58.00  &  32.6$\pm$0.1$^{\prime\prime}$ & 134.7$\pm$0.1$\degr{}$ & --     &   --      &   --      &  $>14.94$       & 14.85$\pm$0.29 &  --            & --             \\
$\upsilon$~And   &  1   &   01:36:49.806 & +41:24:12.36  &  31.5$\pm$0.4$^{\prime\prime}$ & 185.7$\pm$0.8$\degr{}$ & --     &   --      &   --      &  --             & --             & 16.11$\pm$0.35 & $>17.34$       \\
$\upsilon$~And   &  2   &   01:36:46.562 & +41:23:55.39  &  26.5$\pm$0.2$^{\prime\prime}$ & 310.8$\pm$0.1$\degr{}$ & --     &   --      &   --      &   --            & --             & 15.93$\pm$0.22 & --             \\
GJ~581	         &  1   &   15:19:25.370 & -07:43:34.50  &  17.8$\pm$0.2$^{\prime\prime}$ & 312.6$\pm$0.1$\degr{}$ & 14.75	&	14.12	&	13.87	&  13.61$\pm$0.04 & 13.48$\pm$0.05 & n/a            & n/a            \\
GJ~581	         &  2   &   15:19:25.109 & -07:43:13.73  &  17.7$\pm$0.2$^{\prime\prime}$ &  25.7$\pm$0.8$\degr{}$ & --	    &	--   	&	--	    &  16.96$\pm$0.23 & $>18.15$	   & n/a            & n/a            \\
GJ~581           &  3   &   15:19:26.614 & -07:43:44.26  &  23.8$\pm$0.2$^{\prime\prime}$ & 253.9$\pm$0.6$\degr{}$ & --     &   --      &   --      &  $>16.59$       & 16.36$\pm$0.29 & n/a            & n/a            \\
GJ~581           &  4   &   15:19:26.796 & -07:43:45.19  &  25.6$\pm$0.2$^{\prime\prime}$ & 248.6$\pm$0.6$\degr{}$ & --     &   --      &   --      &  --             & $>16.72$       & n/a            & n/a            \\
GJ~849        	 &  1   &   22:09:42.319 & -04:38:25.98  &  21.7$\pm$0.2$^{\prime\prime}$ & 176.5$\pm$0.5$\degr{}$ & --	    &	--   	&	--   	&  15.84$\pm$0.18 & 15.94$\pm$0.24 & n/a            & n/a            \\
GJ~436	         &  1   &   11:42:11.213 & +26:42:32.33  &  16.4$\pm$0.2$^{\prime\prime}$ &  69.9$\pm$0.9$\degr{}$ & --	    &	--	    &	--	    &  16.31$\pm$0.15 & 16.62$\pm$0.24 & n/a            & n/a            \\
HD~69830     	 &  1   &   08:18:22.421 & -12:38:13.85  &  28.0$\pm$0.2$^{\prime\prime}$ & 337.6$\pm$0.2$\degr{}$ & 13.93	&	13.32	&	13.18	&  13.01$\pm$0.11\tablenotemark{*} & 12.83$\pm$0.12\tablenotemark{*} & n/a & n/a \\
\enddata
\tablecomments{(*)The source in the frame of HD69830 is cut off at the edge of the frame; its photometry is estimated.}
\end{deluxetable}

\clearpage
\end{landscape}

\end{document}